\newcommand*{\factor}{0.5}
\documentclass[aps,prc,superscriptaddress,showpacs,nofootinbib,floatfix,preprint]{revtex4-1}

\usepackage{graphicx}
\usepackage{hyperref}
\usepackage{amsmath}
\usepackage{longtable}

\newcommand{\KK}{\ensuremath{K^+K^-}}
\newcommand{\Kmp}{\ensuremath{K^-p}}

\newcommand{\mmp}{\ensuremath{\mbox{MM}_p(\gamma,\phi)}}
\newcommand{\mmd}{\ensuremath{\mbox{MM}_d(\gamma,\phi)}}

\newcommand{\absolute}[1]{\ensuremath{\left| \mathrm{#1} \right|}}
\newcommand{\Eg}{\ensuremath{{E}_{\gamma}}}

\newcommand{\sdmij}{\ensuremath{\rho^{\alpha}_{ij}}}
\newcommand{\sdm}{\ensuremath{\hat{\rho}}}
\newcommand{\Iang}{\ensuremath{W(\Omega,\Phi; \sdm)}}
\newcommand{\Iangi}{\ensuremath{W(\Omega_{i},\Phi_{i}; \sdm)}}
\newcommand{\Obasis}{\ensuremath{O^{lm}}} 

\newcommand{\rhoa}{\ensuremath{\rho^{0}_{00}}}
\newcommand{\rhob}{\ensuremath{\mbox{Re}\rho^{0}_{10}}}
\newcommand{\rhoc}{\ensuremath{\rho^{0}_{1-1}}}
\newcommand{\rhod}{\ensuremath{\rho^{1}_{11}}}
\newcommand{\rhoe}{\ensuremath{\rho^{1}_{00}}}
\newcommand{\rhof}{\ensuremath{\mbox{Re}\rho^{1}_{10}}}
\newcommand{\rhog}{\ensuremath{\rho^{1}_{1-1}}}
\newcommand{\rhoh}{\ensuremath{\mbox{Im}\rho^{2}_{10}}}
\newcommand{\rhoi}{\ensuremath{\mbox{Im}\rho^{2}_{1-1}}}

\newcommand{\phipro}{\ensuremath{\gamma p \to \phi p}}
\newcommand{\phidpn}{\ensuremath{\gamma d \to \phi p n}}
\newcommand{\phidd}{\ensuremath{\gamma d \to \phi d}}

\begin{document}



\title{Measurement of Spin-Density Matrix Elements for $\phi$-Meson
Photoproduction from Protons and Deuterons Near Threshold}


\author{W.C.~Chang}
  \affiliation{Institute of Physics, Academia Sinica, Taipei 11529, Taiwan}

\author{D.S.~Ahn}
  \affiliation{Research Center for Nuclear Physics, Osaka University, Ibaraki, Osaka 567-0047, Japan}
  \affiliation{Department of Physics, Pusan National University, Busan 609-735, Korea}

\author{J.K.~Ahn}
  \affiliation{Department of Physics, Pusan National University, Busan 609-735, Korea}

\author{H.~Akimune}
  \affiliation{Department of Physics, Konan University, Kobe, Hyogo 658-8501, Japan}

\author{Y.~Asano}
  \affiliation{XFEL Project Head Office, RIKEN 1-1, Koto Sayo Hyogo 679-5148, Japan}

\author{S.~Dat\'{e}}
  \affiliation{Japan Synchrotron Radiation Research Institute, Sayo, Hyogo 679-5143, Japan}

\author{H.~Ejiri}
  \affiliation{Research Center for Nuclear Physics, Osaka University, Ibaraki, Osaka 567-0047, Japan}

\author{H.~Fujimura}
  \affiliation{Laboratory of Nuclear Science, Tohoku University, Sendai, Miyagi 982-0826, Japan}
  
\author{M.~Fujiwara}
  \affiliation{Research Center for Nuclear Physics, Osaka University, Ibaraki, Osaka 567-0047, Japan}
  \affiliation{Quantum Beam Science Directorate, Gamma-ray Nondestructive Assay Research Group, Japan Atomic Energy Agency, Tokai-mura, Ibaraki 319-1195, Japan}

\author{S.~Fukui}
  \affiliation{Department of Physics, Nagoya University, Aichi 464-8602, Japan}

\author{S.~Hasegawa}
  \affiliation{Research Center for Nuclear Physics, Osaka University, Ibaraki, Osaka 567-0047, Japan}

\author{K.~Hicks}
  \affiliation{Department of Physics and Astronomy, Ohio University, Athens, Ohio 45701, USA}

\author{K.~Horie}
  \affiliation{Research Center for Nuclear Physics, Osaka University, Ibaraki, Osaka 567-0047, Japan}

\author{T.~Hotta}
  \affiliation{Research Center for Nuclear Physics, Osaka University, Ibaraki, Osaka 567-0047, Japan}

\author{K.~Imai}
  \affiliation{Department of Physics, Kyoto University, Kyoto 606-8502, Japan}

\author{T.~Ishikawa}
  \affiliation{Laboratory of Nuclear Science, Tohoku University, Sendai, Miyagi 982-0826, Japan}

\author{T.~Iwata}
  \affiliation{Department of Physics, Yamagata University, Yamagata 990-8560, Japan}

\author{Y.~Kato}
  \affiliation{Research Center for Nuclear Physics, Osaka University, Ibaraki, Osaka 567-0047, Japan}

\author{H.~Kawai}
  \affiliation{Department of Physics, Chiba University, Chiba 263-8522, Japan}

\author{K.~Kino}
  \affiliation{Research Center for Nuclear Physics, Osaka University, Ibaraki, Osaka 567-0047, Japan}

\author{H.~Kohri}
  \affiliation{Research Center for Nuclear Physics, Osaka University, Ibaraki, Osaka 567-0047, Japan}

\author{N.~Kumagai}
  \affiliation{Japan Synchrotron Radiation Research Institute, Sayo, Hyogo 679-5143, Japan}

\author{P.J.~Lin} 
   \affiliation{Department of Physics, National Kaohsiung Normal University, Kaohsiung 824, Taiwan}

\author{S.~Makino}
  \affiliation{Wakayama Medical University, Wakayama, 641-8509, Japan}

\author{T.~Matsuda}
  \affiliation{Department of Applied Physics, Miyazaki University, Miyazaki 889-2192, Japan}

\author{T.~Matsumura}
  \affiliation{Department of Applied Physics, National Defense Academy in Japan, Yokosuka, Kanagawa 239-8686, Japan}

\author{N.~Matsuoka}
  \affiliation{Research Center for Nuclear Physics, Osaka University, Ibaraki, Osaka 567-0047, Japan}

\author{T.~Mibe}
  \affiliation{Research Center for Nuclear Physics, Osaka University, Ibaraki, Osaka 567-0047, Japan}

\author{M.~Miyabe}
  \affiliation{Department of Physics, Kyoto University, Kyoto 606-8502, Japan}

\author{Y.~Miyachi}
  \affiliation{Department of Physics, Yamagata University, Yamagata 990-8560, Japan}

\author{N.~Muramatsu}
  \affiliation{Research Center for Nuclear Physics, Osaka University, Ibaraki, Osaka 567-0047, Japan}

\author{T.~Nakano}
  \affiliation{Research Center for Nuclear Physics, Osaka University, Ibaraki, Osaka 567-0047, Japan}

\author{M.~Niiyama}
  \affiliation{Department of Physics, Kyoto University, Kyoto 606-8502, Japan}

\author{M.~Nomachi}
  \affiliation{Department of Physics, Osaka University, Toyonaka, Osaka 560-0043, Japan}

\author{Y.~Ohashi}
  \affiliation{Japan Synchrotron Radiation Research Institute, Sayo, Hyogo 679-5143, Japan}

\author{H.~Ohkuma}
  \affiliation{Japan Synchrotron Radiation Research Institute, Sayo, Hyogo 679-5143, Japan}

\author{T.~Ooba}
  \affiliation{Department of Physics, Chiba University, Chiba 263-8522, Japan}

\author{D.S.~Oshuev}
  \affiliation{Institute of Physics, Academia Sinica, Taipei 11529, Taiwan}

\author{C.~Rangacharyulu}
  \affiliation{Department of Physics and Engineering Physics, University of Saskatchewan, Saskatoon SK S7N 5E2, Canada}

\author{A.~Sakaguchi}
  \affiliation{Department of Physics, Osaka University, Toyonaka, Osaka 560-0043, Japan}

\author{P.M.~Shagin}
  \affiliation{School of Physics and Astronomy, University of Minnesota, Minneapolis, Minnesota 55455, USA}

\author{Y.~Shiino}
  \affiliation{Department of Physics, Chiba University, Chiba 263-8522, Japan}

\author{H.~Shimizu}
  \affiliation{Laboratory of Nuclear Science, Tohoku University, Sendai, Miyagi 982-0826, Japan}

\author{Y.~Sugaya}
  \affiliation{Department of Physics, Osaka University, Toyonaka, Osaka 560-0043, Japan}

\author{M.~Sumihama}
  \affiliation{Research Center for Nuclear Physics, Osaka University, Ibaraki, Osaka 567-0047, Japan}

\author{Y.~Toi}
  \affiliation{Department of Applied Physics, Miyazaki University, Miyazaki 889-2192, Japan}

\author{H.~Toyokawa}
  \affiliation{Japan Synchrotron Radiation Research Institute, Sayo, Hyogo 679-5143, Japan}

\author{M.~Uchida}
  \affiliation{Department of Physics, Tokyo Institute of Technology, Tokyo 152-8551, Japan}

\author{A.~Wakai}
  \affiliation{Akita Research Institute of Brain and Blood Vessels, Akita, 010-0874, Japan}

\author{C.W.~Wang}
  \affiliation{Institute of Physics, Academia Sinica, Taipei 11529, Taiwan}

\author{S.C.~Wang}
  \affiliation{Institute of Physics, Academia Sinica, Taipei 11529, Taiwan}

\author{K.~Yonehara}
  \affiliation{Department of Physics, Konan University, Kobe, Hyogo 658-8501, Japan}

\author{T.~Yorita}
  \affiliation{Research Center for Nuclear Physics, Osaka University, Ibaraki, Osaka 567-0047, Japan}
  \affiliation{Japan Synchrotron Radiation Research Institute, Sayo, Hyogo 679-5143, Japan}

\author{M.~Yoshimura}
  \affiliation{Institute for Protein Research, Osaka University, Suita, Osaka, 565-0871, Japan}

\author{M.~Yosoi}
  \affiliation{Research Center for Nuclear Physics, Osaka University, Ibaraki, Osaka 567-0047, Japan}

\author{R.G.T.~Zegers} 
   \affiliation{National Superconducting Cyclotron Laboratory, Michigan State University, East Lansing, Michigan 48824, USA}

\collaboration{LEPS Collaboration}
\noaffiliation



\begin{abstract}

The LEPS/SPring-8 experiment made a comprehensive measurement of the
spin-density matrix elements for \phipro, \phidpn~ and \phidd~ at
forward production angles. A linearly polarized photon beam at
$\Eg=1.6-2.4$ GeV was used for the production of $\phi$ mesons.  The
natural-parity Pomeron exchange processes remain dominant near
threshold. The unnatural-parity processes of pseudoscalar exchange is
visible in the production from the nucleons but is greatly reduced in
the coherent production from deuterons. There is no strong \Eg\
dependence, but there is some dependence on momentum transfer. A small
but finite value of the spin-density matrix elements reflecting
helicity-nonconserving amplitudes in the $t$-channel is observed.

\end{abstract}


\maketitle


\section{Introduction}
\label{sec:intro}

The study of photoproduction of light vector mesons ($\rho$, $\omega$,
and $\phi$) plays an important role in hadron physics to understand
the nonperturbative aspect of QCD. At high energies ($W>10$ GeV),
Pomeron exchange in the $t$ channel describes well the photoproduction
mechanism in the framework of vector-meson dominance. The Pomeron is
introduced in the Regge theory for describing high-energy hadron
scattering, and it is generally believed to originate from multi gluon
exchange processes. At low energies near threshold ($W \sim 2$ GeV),
the reactions are open to the exchange of mesons and baryons, where
possible ``missing resonances'' could be involved. In particular, the
gluon dynamics dominates in the $\phi$-meson production process
because the quark exchange is suppressed by the Okubo-Zweig-Iizuka
(OZI) rule owing to the major $s\bar{s}$ content of the $\phi$
meson. Therefore, the photoproduction of $\phi$ mesons at low energies
provides a unique opportunity to explore the behavior of Pomeron
exchange and to study exotic hadronic interactions mediated by multi
gluon exchanges. In fact, $\phi$-meson photoproduction from protons
and deuterons has been measured in the Laser Electron Photon beamline
at SPring-8
(LEPS)~\cite{LEPS_phip,LEPS_phiD_co,LEPS_phiD_inco,LEPS_phiA_inco},
the CEBAF Large Acceptance Spectrometer (CLAS) at Jefferson
Lab~\cite{CLAS_phip,CLAS_phiD_co,CLAS_phiD_inco}, and the Spectrometer
Arrangement for Photon induced Reactions (SAPHIR) at
Bonn~\cite{SAPHIR_phip}.

Spin observables are known to be a powerful tool to obtain further
insights into the relevant reaction mechanisms. With the use of
linearly polarized photons, the decay angular distribution of the
vector meson $\phi$ can be expressed in terms of the real or imaginary
part of nine spin-density matrix elements
(\sdmij)~\cite{Schilling}. These matrix elements are sensitive to the
underlying reaction
mechanism~\cite{Titov_iso,Titov_iso2,Titov_spinamp,Titov_coherent,Titov_coherent2}. For
example, the matrix element \rhog\ reflects the asymmetry in the
contribution from natural-parity and unnatural-parity
exchange~\cite{Titov_spinamp}. There were some measurements of decay
asymmetry~\cite{phi_Hpalpern} and spin-density matrix
elements~\cite{phi_Ballam,phi_Atkinson} of $\phi$ production from free
protons at high energies (2.8, 4.7, 9.3 and 20$-$40 GeV). The results
were consistent with $s$-channel helicity conservation. Judging from
the fact that $\rhog \approx 0.5$, one can conclude that the $\phi$
meson is produced predominately by natural-parity Pomeron exchange at
high energies.

At $E_{\gamma}=1.6-2.3$ GeV, the photoproduction of $\phi$ mesons from
free protons in the forward direction~\cite{LEPS_phip} can be mostly
described by the Pomeron and ($\pi,\eta$) exchange in the $t$ channel.
A value of 0.2 was found for \rhog, which deviates from 0.5, the limit
corresponding to pure natural-parity exchange. It suggests a
nonnegligible contribution of unnatural-parity exchange processes, in
contrast to complete dominance of the Pomeron exchange at high
energies.

In the measurement of coherent $\phi$ production from
deuterons~\cite{LEPS_phiD_co}, it is reported that \rhog\ clearly
becomes close to 0.5. This suggests that the dominant unnatural-parity
component, isovector $\pi$ exchange, is forbidden in the coupling to
the isoscalar deuteron target~\cite{Titov_coherent,Titov_coherent2}.

The nuclear transparency ratio of $\phi$ photoproduction for deuterons
shows a large suppression of incoherent
production~\cite{LEPS_phiD_inco} and is consistent with the $A$
dependence of the ratio observed for nuclear
targets~\cite{LEPS_phiA_inco}. Also, \rhog\ was observed to be
slightly larger in incoherent production from deuterons, compared with
production from free protons. It coincides with destructive
interference between isovector $\pi$ and isoscalar $\eta$ exchange
amplitudes in the $\gamma n \to \phi n$
reaction~\cite{Titov_iso,Titov_iso2}. However, there is no observation
of a large isospin asymmetry of $\phi$ production from nucleons. The
decrease in the $\phi$-meson yields, scaled as number of nucleons in
the production from deuterons, cannot be adequately explained in terms
of isospin asymmetry.

A narrow bump structure was found around $E_{\gamma}=2$ GeV in
$\phi$-meson production cross sections from nucleons~\cite{LEPS_phip}
and the origin of this structure is not yet understood. Naively, it is
speculated to result from the appearance of ($\pi,\eta$) exchange near
threshold. However, the measured decay angular distributions, which in
principle reflect the relative weights of natural-parity and
unnatural-parity processes, do not vary appreciably across the bump
region. This structure cannot be described by a conventional model
with the Pomeron and pseudoscalar exchange, where a monotonic energy
dependence is predicted~\cite{LEPS_phip}. Various theoretical
interpretations of this structure are proposed: the interference
between unnatural-parity $\pi$ and $\eta$
exchanges~\cite{Titov_coherent2}, the interference of the isovector
scalar $a_{0}$ meson with Pomeron exchange~\cite{Titov_coherent2}, the
coupled-channel effect of $K\Lambda(1520)$ and $\phi
N$~\cite{Hosaka_phi}, and the existence of $N^{*}$ resonances with a
large $s\bar{s}$ component~\cite{Hosaka_phi,
Kiswandhi_phi}. Nonetheless, a satisfactory description of the bump
structure of $\phi$-meson photoproduction is still
lacking. Comprehensive information on spin-density matrix elements
will help clarify the situation.

Previously, only limited numbers of spin-density matrix elements were
determined from one-dimensional decay angular
distributions~\cite{LEPS_phip, LEPS_phiD_co, LEPS_phiD_inco,
CLAS_phiD_co, CLAS_phiD_inco, SAPHIR_phip}. This paper presents the
results of a complete set of spin-density matrix elements in the
\phipro, \phidpn, and \phidd\ reactions measured in the LEPS/SPring-8
experiment. The angular distributions were measured via the
charged-kaon decay mode of the $\phi$ meson. In
Section~\ref{sec:method} we describe the formulation of spin-density
matrix elements and the extended maximum likelihood fit used for the
determination. The experimental setup and the analysis details are
introduced in Sec.~\ref{sec:exp}. Sec.~\ref{sec:results} shows the
spin-density matrix elements of photoproduction of $\phi$ mesons from
protons and deuterons. The interpretation of the data and discussion
are given in Sec.~\ref{sec:discussion}. Finally,
Sec.~\ref{sec:summary} provides a summary.

\section{Spin-density Matrix Elements and Extended Maximum Likelihood}
\label{sec:method}

\subsection{Decay angular distribution}
\label{subsec:angular}

The angular distribution, $W$, of $K^+$ decaying from $\phi$
mesons produced with a linearly polarized photon beam can be expressed
as follows~\cite{Schilling}:
\begin{equation}
\label{eq:decayang1}
\begin{split}
  W(\Omega,\Phi;\hat{\rho})= & W(\cos \theta,\varphi,\Phi;\hat{\rho}) \\ 
  =& W^0(\cos \theta, \varphi;\hat{\rho}) \\ 
   & - P_{\gamma} \cos2\Phi W^1(\cos\theta, \varphi;\hat{\rho}) \\ 
   & - P_{\gamma} \sin2\Phi W^2(\cos \theta, \varphi;\hat{\rho})
\end{split}
\end{equation}

\begin{eqnarray}
\label{eq:decayang2}
W^0&=&\frac{3}{4\pi}[ \frac{1}{2}(1-\rhoa) +
	  \frac{1}{2}(3\rhoa-1)\cos^2\theta \nonumber \\ &&
-\sqrt{2}\rhob \sin2\theta \cos\varphi
	  -\rhoc \sin^2\theta \cos2\varphi], \\ 
W^1&=&\frac{3}{4\pi}[\rhod \sin^2\theta
	  +\rhoe \cos^2\theta \nonumber \\ &&
-\sqrt{2}\rhof\sin2\theta \cos\varphi -\rhog
	  \sin^2\theta \cos2\varphi], \\ 
W^2&=&\frac{3}{4\pi}[\sqrt{2}\rhoh \sin2\theta \sin\varphi \nonumber \\ &&
	  +\rhoi\sin^2\theta \sin2\varphi]
\end{eqnarray}
Here, $\hat{\rho}$ represents the measurable parts of the nine
independent spin-density matrix elements \sdmij, $\theta$ and
$\varphi$ are the polar and azimuthal angles of decay particles in the
rest frame of vector mesons, $\Phi$ is the azimuthal angle of the
photon electric polarization vector with respect to the production
plane of vector mesons, and $P_{\gamma}$ is the degree of linear
polarization of incident photons.

Conventionally, there are three different choices of the quantization
axis $z^\prime$ for the decay reference system~\cite{Schilling}: the
{\em helicity} system with $z^\prime$ opposite to the velocity of the
recoiling nucleons in the vector-meson rest frame, the
Gottfried-Jackson system with $z^\prime$ parallel to the momentum of
the photon in the vector-meson rest frame, and the Adair system with
$z^\prime$ parallel to the photon momentum in the overall
center-of-mass (CM) system.

The physical property of the production mechanism is simplest when
illustrated in: the helicity system for $s$-channel helicity
conservation, the Gottfried-Jackson system for $t$-channel helicity
conservation with no absorption, and the Adair system for the spin
independence in the overall CM system~\cite{phi_Ballam2}. In addition,
the Gottfried-Jackson system has the advantage that some spin-density
matrix elements work as a measure of the asymmetry between processes
with natural-parity and unnatural-parity exchanges in the $t$
channel~\cite{Titov_iso2}. In general, the values of spin-density
matrix elements depend on the system chosen. Nevertheless, the
difference among the three systems becomes small at very forward
angles. The results in the following are presented in all three
systems for easy comparison.

The contribution to the cross section from natural-parity and
unnatural-parity exchanges in the $t$ channel, $\sigma^N$ and
$\sigma^U$, can be determined from the density matrix elements. In the
case of helicity-conserving exchanges, all spin-density matrix
elements become zero, except for \rhog\ and
\rhoi~\cite{Titov_spinamp}. That $\rhog=-\rhoi=+0.5$ ($-0.5$)
corresponds to the case with pure natural-parity (unnatural-parity)
exchange. The matrix element \rhog\ directly relates to the asymmetry
in the contributions from natural-parity and unnatural-parity
exchanges~\cite{Titov_spinamp} and is expressed as
\begin{equation}
\rho^1_{1 -1}=\frac{1}{2} \frac{\sigma^N-\sigma^U}{\sigma^N+\sigma^U}.
\label{eq:nunratio}
\end{equation}
In principle, nonzero values for spin-density matrix elements other
than \rhog\ and \rhoi\ indicate the nonconventional OZI-evading
processes such as the $s\bar{s}$ knockout~\cite{Titov_ssknockout} or
nondiffractive baryon resonance
production~\cite{Titov_spinamp,Zhao}. Those processes are commonly
expected to become important at large transferred-momentum ($|t|$)
regions for $\phi$ photoproduction.

The matrix element $\rho^{0}_{00}$ reflects the strength of the
single-spin-flip exchange amplitude resulting from the components
other than the Pomeron exchange. For evaluation of the relative
contribution of the unnatural-parity exchange at small $|t|$, one need
take into account \rhoa~\cite{Titov_spinamp} as follows:
\begin{equation}
\frac{\sigma^U}{\sigma^N+\sigma^U+\sigma^{spin-flip}}=\frac{1}{2}(1-2\rhog-\rhoa).
\end{equation}

A finite value of the spin-density matrix element \rhoc\ comes from
the amplitude responsible for the double-spin transitions where the
helicity of the $\phi$ meson differs from the photon helicity by two
units: $\lambda_{\gamma} = \pm 1 \to \lambda_{\phi} = \mp1 $. In many
theoretical models for $\phi$-meson photoproduction, such as scalar,
pseudoscalar $t$-channel exchange, and the original
Donnachie-Landshoff Pomeron exchange model based on the
Pomeron-isoscalar photon identity, these transitions are forbidden,
and \rhoc\ is exactly zero. In the modified Donnachie-Landshoff model,
\rhoc\ in the Gottfried-Jackson system could be nonzero owing to the
spin-orbital interaction inherent to the two-gluon exchange in the
$t$ channel~\cite{Titov_iso2,Laget}.


\subsection{Extended maximum likelihood fit}
\label{subsec:mlf}

Experimentally we measure the angular distribution \Iang\ for charged
kaons decaying from photoproduction events of the $\phi$ meson. A few
spin-density matrix elements can be extracted from the one-dimensional
angular distributions of $\cos\theta$, $\cos\varphi$, $\cos\Phi$, and
$\cos(\varphi-\Phi)$~\cite{Titov_spinamp}. In this work, a binless
maximum likelihood fit is applied to determine the nine spin-density
matrix elements simultaneously.

A likelihood function with a perfect detection efficiency, $\cal L$,
is defined as:
\begin{equation}
\begin{split}
{\cal L} & =  \prod_{i=1}^{N} P_{i} \\
& =  \prod_{i=1}^{N} \frac{\Iangi} {\int \Iang d\Omega d\Phi }
\end{split}
\end{equation}
with $P_{i}$ the normalized likelihood for each event and \Iang\ the
probability density function. The logarithmic likelihood function $-
\ln {\cal L}$ is minimized by the CERNLIB MINUIT package~\cite{MINUIT}.

If the detector efficiency is not perfect, the probability density
function is modified as
\begin{equation}
P_i = \frac{\Iangi}{\int \Iang \eta(\Omega) d\Omega d\Phi} =
\frac{\Iangi}{{\cal{W}} (\hat{\rho})}.
\end{equation}
By taking into account the efficiency $\eta(\Omega)$, the evaluation of
the normalization factor $\cal{W}$ with a variation of $\hat{\rho}$ in
the minimizing process, is technically challenging. This issue can be
solved by representing the angular distribution \Iang\ with a set of
orthogonal bases $O^{lm}$~\cite{Obasis_Chung}:
\begin{equation}
\Iang = \sqrt{\frac{3}{4\pi}} \sum_{l,m}a_{lm}(\hat{\rho})O^{lm}(\Omega,\Phi),
\end{equation}
where the orthogonal conditions for each base are satisfied:
\begin{equation}
	\int O^{*lm} O^{l^\prime m^\prime}~d\Omega~d\Phi
= \delta_{ll^\prime} \delta_{mm^\prime},
\end{equation}
and the coefficient for each basis $a_{lm}$ is
\begin{equation}
	\int O^{*lm} \Iang ~d\Omega~d\Phi =
\sqrt{\frac{3}{4\pi}}a_{lm}(\hat{\rho})
\end{equation}
The explicit representations of $O^{lm}$ in terms of spherical
harmonics, and the coefficients $a_{lm}$'s are given in
Table~\ref{tab:obasis}.

\begin{table}
\caption{Angular coefficients $a_{lm}$ and their corresponding bases $O^{lm}$.}
\begin{tabular}[t]{ccc} \hline \hline
l,m &
$a_{lm}$ &
$O^{lm}$ \\ \hline
0,1 & 
$\sqrt{2\pi}\sqrt{\frac{1}{3}}$ & 
$Y_0^0(\frac{1}{\sqrt{2\pi}})$ \\
0,2 &
$\sqrt{2\pi}\sqrt{\frac{1}{15}}(3\rho^{0}_{00}-1)$ &
$Y_2^{0\ast}(\frac{1}{\sqrt{2\pi}})$ \\
0,3 &
$-\sqrt{2\pi}\sqrt{\frac{8}{5}}{\rm Re}(\rho^{0}_{10})$ &
$\frac{1}{\sqrt{2}}(-Y_2^{1\ast} + Y_2^{-1\ast})(\frac{1}{\sqrt{2\pi}})$ \\
0,4 &
$-\sqrt{2\pi}\sqrt{\frac{4}{5}}\rho^{0}_{1-1}$ &
$\frac{1}{\sqrt{2}}(Y_2^{2\ast} + Y_2^{-2\ast})(\frac{1}{\sqrt{2\pi}})$ \\
1,1 &
$-\sqrt{\pi}P_{\gamma}\sqrt{\frac{1}{3}}(\rho^{1}_{00}+2\rho^{1}_{11})$ &
$Y_0^0(\frac{1}{\sqrt{\pi}}\cos(2\Phi))$ \\
1,2 &
$-\sqrt{\pi}P_{\gamma}\sqrt{\frac{4}{15}}(\rho^{1}_{00}-\rho^{1}_{11})$ &
$Y_2^{0\ast}(\frac{1}{\sqrt{\pi}}\cos(2\Phi))$ \\
1,3 &
$\sqrt{\pi}P_{\gamma}\sqrt{\frac{8}{5}}\rho^{1}_{10}$ &
$\frac{1}{\sqrt{2}}(-Y_2^{1\ast} + Y_2^{-1\ast})(\frac{1}{\sqrt{\pi}}\cos(2\Phi))$ \\
1,4 &
$\sqrt{\pi}P_{\gamma}\sqrt{\frac{4}{5}}\rho^{1}_{1-1}$ &
$\frac{1}{\sqrt{2}}(Y_2^{2\ast} + Y_2^{-2\ast})(\frac{1}{\sqrt{\pi}}\cos(2\Phi))$ \\
2,1 &
$-\sqrt{\pi}P_{\gamma}\sqrt{\frac{8}{5}}{\rm Im}(\rho^{2}_{10})$ &
$\frac{1}{i}\frac{1}{\sqrt{2}}(Y_2^{1\ast} + Y_2^{-1\ast})(\frac{1}{\sqrt{\pi}}\sin(2\Phi))$ \\
2,2 &
$-\sqrt{\pi}P_{\gamma}\sqrt{\frac{4}{5}}{\rm Im}(\rho^{2}_{1-1})$ &
$\frac{1}{i}\frac{1}{\sqrt{2}}(Y_2^{2\ast} - Y_2^{-2\ast})(\frac{1}{\sqrt{\pi}}\sin(2\Phi))$ \\ \hline
\end{tabular}
\label{tab:obasis}
\end{table}


The angular distribution of the detector efficiency $\eta(\Omega)$ can
also be expanded by the same orthogonal basis \Obasis:
\begin{equation}
\eta(\Omega)=\sum_{l,m}b_{lm}O^{lm}(\Omega).
\label{eq:blm}
\end{equation}
In a kinematic bin, the angular moment of experimental acceptance,
$b_{lm}$, is evaluated as
\begin{eqnarray}
    b_{lm} &=& \sqrt{\frac{3}{4\pi}}
    \int O_{lm}(\Omega,~\Phi) \eta (\Omega) d\Omega d\Phi \\ 
    &=& \frac{1}{N_{\rm generated}} \sum_{i=1}^{N_{\rm accepted}}
    O^{lm}(\Omega_{i},\Phi_{i}),
\end{eqnarray}
where the numerator and denominator sum over the accepted and
isotropically generated Monte Carlo (MC) events, respectively.

The evaluation of the normalization factor $\cal{W}(\hat{\rho})$ is
simplified as follows:
\begin{eqnarray}
    \cal{W}(\hat{\rho}) &=& \int \Iang  \eta(\Omega) d\Omega d\Phi \\
    &=& \sum_{l,m} a_{lm}(\hat{\rho}) b_{lm}.
\end{eqnarray}
Furthermore, the restriction that the probability distribution $P$ is
normalized to 1 could be relaxed in an ``extended maximum likelihood
fit''~\cite{EML_Lyons}. The integral of the unnormalized probability
density function $\cal{P}$ represents the total number of predicted
events, $\bar{N}$, under the assumption of a Poisson variation for the
measured number of events, $N$.

The extended likelihood function can be written as
\begin{eqnarray}
{\cal{L}} &=& \Biggl(\frac{\bar{N}^{N}e^{-\bar{N}}}{N!}\Biggr)\prod_{i=1}^N{\cal{P}}_{i} \\
{\cal{P}}_{i}&=&\frac{ {\cal{Y}} \Iangi} {\int{} {\cal{Y}} \Iang \eta(\Omega) d\Omega d\Phi} \\
\bar{N} &=& {\int{} {\cal{Y}} \Iang \eta(\Omega) d\Omega d\Phi} = {\cal{Y}}{\cal{W}}(\hat{\rho}),
\end{eqnarray}
where $\cal{Y}$ is the actual yield. By neglecting the terms that do not
depend on the fit parameters, the likelihood function to be minimized
can be rewritten as follows:
\begin{eqnarray}
-\ln {\cal{L}} (\hat{\rho},\bar{N}) & = & - \sum_{i=1}^N \ln [{\cal{Y}} I(\Omega_{i};\hat{\rho})] + \bar{N} \\
& = & - \sum_{i=1}^N \ln [\bar{N} I(\Omega_{i};\hat{\rho})/ {\cal{W}}(\hat{\rho}))] + \bar{N}.
\label{eq:likelohood}
\end{eqnarray}
Now besides $\hat{\rho}$, the expected number of events, $\bar{N}$, becomes an additional parameter in the minimization, and it should
turn out to be the number of events to be fit, $N$, in the final
result. This fact could be utilized to validate the success of the
minimization procedure.

\section{Experiment and Analysis}
\label{sec:exp}

The experiment was carried out at the LEPS/SPring-8 facility using a
linearly polarized photon beam produced by backward Compton scattering
of the Ar laser from 8-GeV electrons in the storage ring of
SPring-8. Photons in the energy range of 1.5$-$2.4 GeV were tagged by
detecting recoil electrons. The photon beam with an intensity of
$\sim$10$^6$/s was directed onto liquid hydrogen or deuterium
targets inside a 15-cm-long target cell. The direction of linear
polarization was controlled vertically or horizontally by using a
half-wave plate for the laser with a polarization of nearly
100\%. Charged particles emitted from the interaction points of
photons with a target were detected at forward angles in the LEPS
spectrometer. The spectrometer consisted of a start counter, a
silica-aerogel \v{C}erenkov counter, a silicon vertex detector, a
dipole magnet, three multiwire drift chambers, and a time-of-flight
wall. The angular coverage of the spectrometer was about 0.4 rad and
0.2 rad in the horizontal and vertical directions,
respectively. Particle identification was made by mass reconstruction
using the measured time of flight and momentum. For more details
concerning the detector configuration and the quality of particle
identification, see Ref.~\cite{LEPS}. In the present work, the
integrated number of tagged photons reached at 5.6$\times$10$^{12}$
(4.6$\times$10$^{12}$) for the hydrogen (deuterium) runs.

\begin{figure}[htbp]
\includegraphics[width=\factor\textwidth]{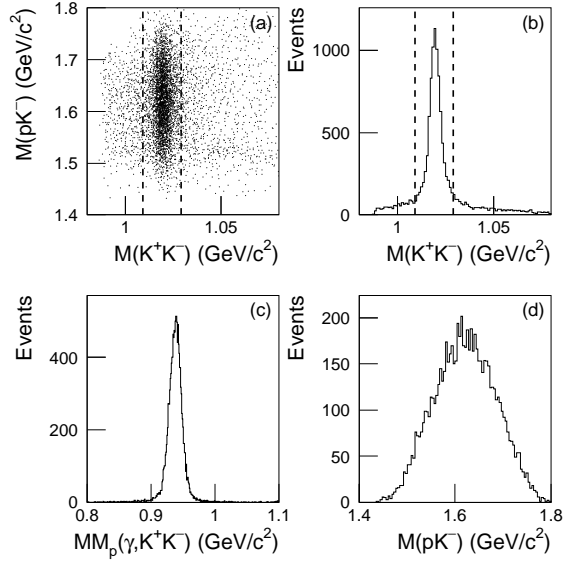}
\caption{Spectra for the $p(\gamma, \KK)X$ reaction within the
experimental acceptance: (a) Dalitz plot of $\KK p$ final state, (b)
the invariant mass spectrum of \KK, (c) missing mass spectrum, and (d)
the invariant mass spectrum of \Kmp\ for the selected $\phi$
events. The dashed lines on the invariant mass of \KK\ in (a) and (b)
label the region for the selection of $\phi$ events.}
\label{fig:spectrum1_LH2}
\end{figure}

\begin{figure}[htbp]
\includegraphics[width=\factor\textwidth]{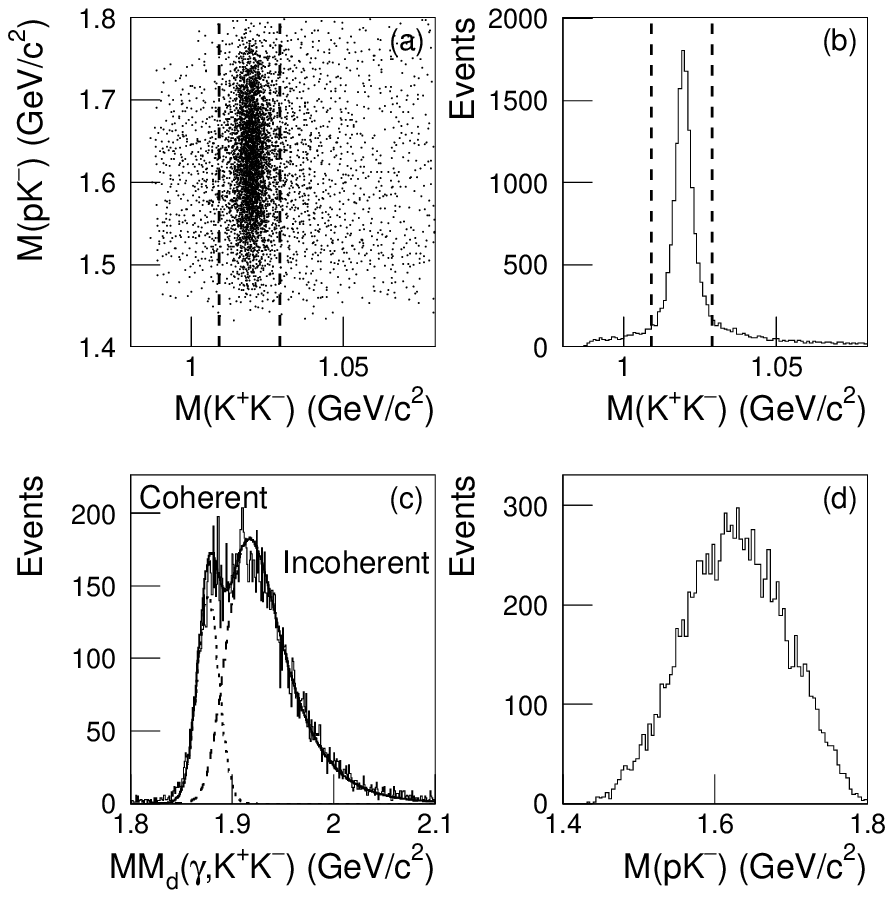}
\caption{Spectra for the $d(\gamma, \KK)X$ reaction within the
experimental acceptance: (a) Dalitz plot of $\KK p$ final state, (b)
the invariant mass spectrum of \KK, (c) missing mass spectrum assuming
rest deuteron target, and (d) the invariant mass spectrum of \Kmp\
assuming rest proton target for the selected $\phi$ events. The dashed
lines on the invariant mass of \KK\ in (a) and (b) label the region
for the selection of $\phi$ events. The $\mmd$ spectrum in (c) is
fitted with the sum (solid line) of MC-simulated components of
coherent (dotted line) and incoherent (dashed line) events.}
\label{fig:spectrum1_LD2}
\end{figure}

Events with both $K^+$ and $K^-$ tracks detected were selected. The
spectra of a Dalitz plot of the $\KK p$ final state, \KK\ invariant
mass, missing mass and \Kmp\ invariant mass for those events from
hydrogen and deuterium targets are shown Fig.~\ref{fig:spectrum1_LH2}
and Fig.~\ref{fig:spectrum1_LD2}, respectively. A clear $\phi$ peak in
the nominal mass of $\phi$ meson, 1.019 GeV, is seen in
Fig.~\ref{fig:spectrum1_LH2}(b) and
Fig.~\ref{fig:spectrum1_LD2}(b). The missing mass spectrum of the
events from hydrogen shows a peak around the proton mass as seen in
Fig.~\ref{fig:spectrum1_LH2}(c) while more complex structure is
observed in the spectrum of the production from deuterium (see
Fig.~\ref{fig:spectrum1_LD2}(c)). In the missing mass spectrum,
assuming the whole deuteron as the target ($\mmd$), events of coherent
$\phi$ production, \phidd, peak at the deuteron mass of 1.875
GeV/c$^2$ whereas incoherent events, \phidpn, are distributed at
relatively higher mass. This missing-mass spectrum is nicely
reproduced by MC simulations of coherent and incoherent
$\phi$ production processes. The MC simulation takes into account
experimental parameters such as geometrical acceptance, energy and
momentum resolutions, and the efficiency of detectors. The effects of
Fermi motion, along with off-shell aspects of target nucleons inside
deuterium and final-state interaction between the target and
spectator nucleons, are also included to describe the $\mmd$
distribution of incoherent events~\cite{LEPS_phiD_co,LEPS_phiD_inco}.

Furthermore, $\phi$-meson events were singled out with a cut on the
invariant mass of a \KK\ pair $|{\rm M}(K^+K^-)-{\rm M}_{\phi}|<$0.01
GeV/c$^2$, which is indicated as the region between two dashed lines
in Fig.~\ref{fig:spectrum1_LH2}(b) and
Fig.~\ref{fig:spectrum1_LD2}(b). For the \phipro\ events, an
additional cut on the missing mass $\absolute{\mmp}<$0.03 GeV/c$^2$ is
required. As seen in Figs.~\ref{fig:spectrum1_LH2}(a) and
\ref{fig:spectrum1_LH2}(d) and Figs.~\ref{fig:spectrum1_LD2}(a) and
\ref{fig:spectrum1_LD2}(d), the contamination of $\Lambda(1520)$ in
the selected $\phi$-meson events is insignificant.

\begin{figure}[htbp]
\includegraphics[width=\factor\textwidth]{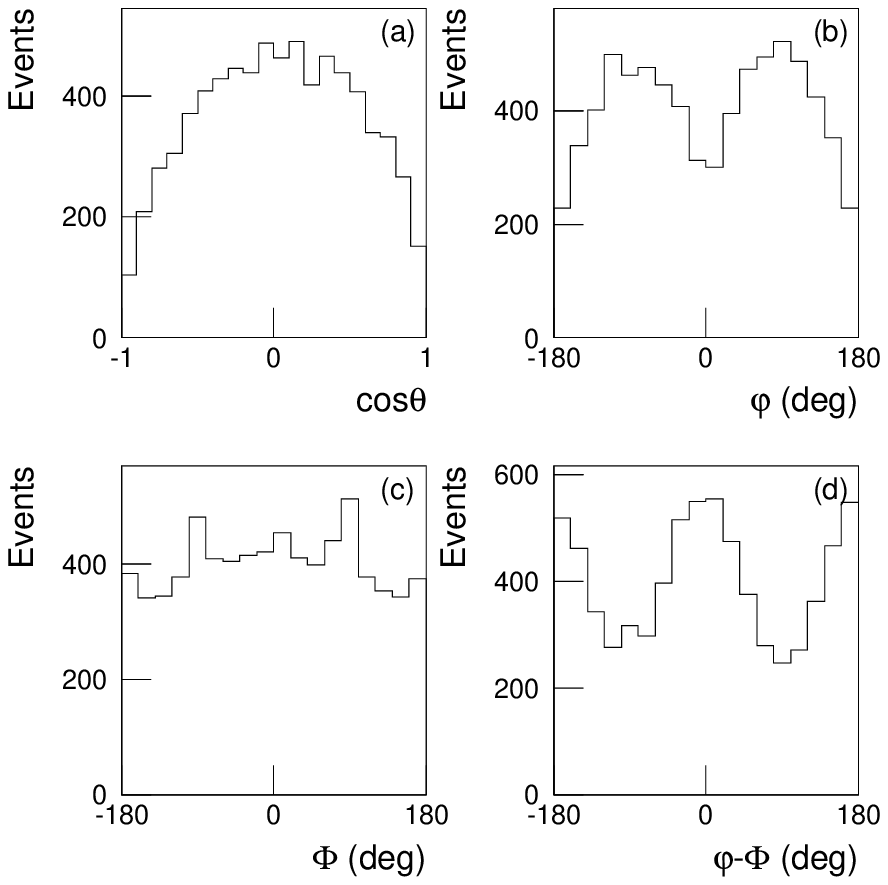}
\caption{Decay angular distributions for the $p(\gamma, \KK)X$ reaction in the
helicity system within the experimental acceptance: (a)
$W(\cos\theta)$, (b) $W(\varphi)$, (c) $W(\Phi)$, and (d)
$W(\varphi-\Phi)$.}
\label{fig:spectrum2_LH2}
\end{figure}

\begin{figure}[htbp]
\includegraphics[width=\factor\textwidth]{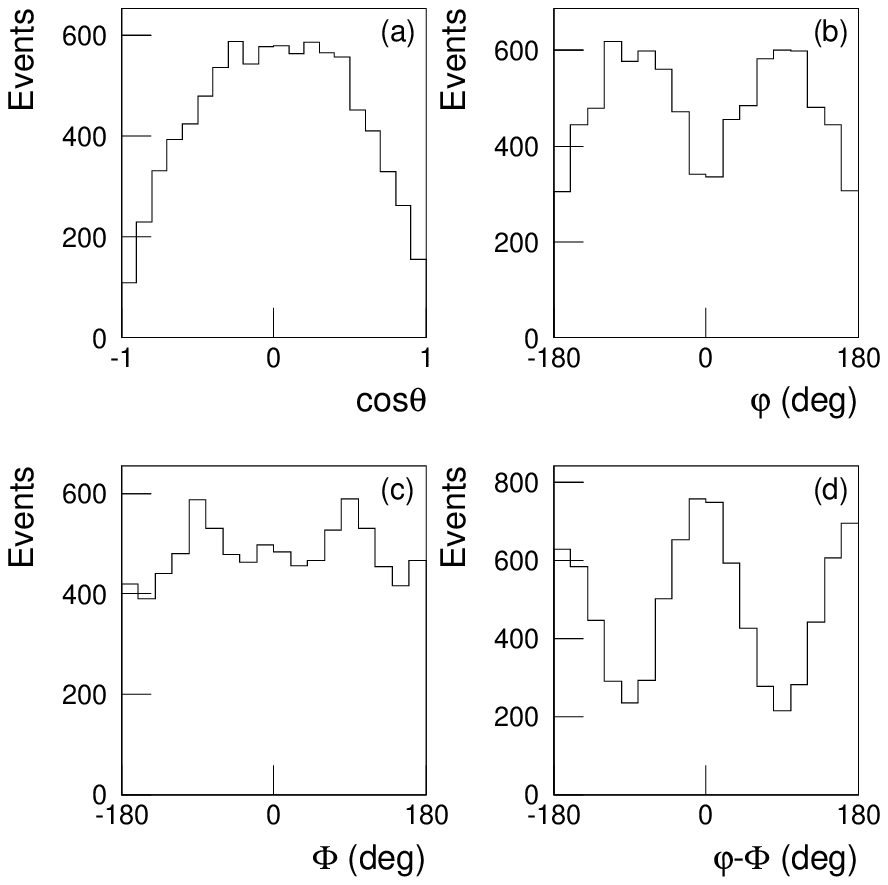}
\caption{Decay angular distributions for the $d(\gamma, \KK)X$ reaction in the
helicity system within the experimental acceptance: (a)
$W(\cos\theta)$, (b) $W(\varphi)$, (c) $W(\Phi)$, and (d)
$W(\varphi-\Phi)$.}
\label{fig:spectrum2_LD2}
\end{figure}

The decay angular distributions of $W(\cos\theta)$, $W(\varphi)$,
$W(\Phi)$, and $W(\varphi-\Phi)$ in the helicity system for $\phi$
events within the experimental acceptance are shown in
Fig.~\ref{fig:spectrum2_LH2} and Fig.~\ref{fig:spectrum2_LD2}. The
polar angle distribution $W(\cos\theta)$ behaves as $\sim$$
(3/4)\sin^2\theta$, indicating the dominance of
$s$-channel-helicity-conserving processes. That the azimuthal angle
$\varphi$ of decay daughter $K^+$ aligns along that of the electric
polarization vector $\Phi$ of the incident photon signifies a larger
contribution from natural-parity exchange processes. Such a
correlation between the two angles, $\varphi$ and $\Phi$, is seen to
be even stronger in the production from deuterium. Quantitative
information will be given in Sec.~\ref{sec:results}.

Since not all particles in the final state are detected, there is a
mixture of two components, incoherent and coherent, in the selected
$\phi$ events from deuterium. To disentangle the individual
contributions, extra effort is needed. By following the prescription
specified in Refs.~\cite{LEPS_phiD_co,LEPS_phiD_inco}, the percentage
of incoherent and coherent events, ${\cal{R}}^1$ and ${\cal{R}}^2$,
are determined in the two separated missing mass regions,
respectively, by a fit of the missing mass spectrum \mmd\ with the
MC-simulated distributions. The division of these two regions, $\rm
MM_{\rm div}$, is chosen to be 1.89 GeV/c$^2$. Event by event,
relative weights composed of either ${\cal{R}}^1$ or ${\cal{R}}^2$ are
assigned to the likelihood of incoherent and coherent processes,
depending on where the missing-mass of the event sits. In contrast to
Eq.~(\ref{eq:likelohood}), the likelihood function for each event is
represented as the weighted sum of the individual likelihoods from the
incoherent and coherent processes:
\begin{equation}
\begin{split}
-\ln {\cal{L}} (\hat{\rho}^{\rm inco},\hat{\rho}^{\rm co},\bar{N}) =&  \\
-\sum_{i=1}^N \ln
\{ \bar{N}[{\cal{R}}_i\frac{W(\Omega_{i};\hat{\rho}^{\rm inco})}{{\cal{W}}(\hat{\rho}^{\rm inco})}
&+(1-{\cal{R}}_i)\frac{W(\Omega_{i};\hat{\rho}^{\rm co})}{{\cal{W}}(\hat{\rho}^{\rm co})}]\} +\bar{N}.
\end{split}
\end{equation}
Here ${\cal{R}}_i$ is either ${\cal{R}}^1$ or ${\cal{R}}^2$ according
to the associated missing-mass value of the event; the spin-density
matrix elements for the incoherent reaction $\hat{\rho}^{\rm inco}$
and coherent reaction $\hat{\rho}^{\rm co}$, together with the
expected number of events, $\bar{N}$, are the parameters to be
determined in the extended maximum likelihood fit.

The extended maximum likelihood fit was performed in the framework of
the MINUIT package. To locate the global minimum, the initial values
of fit parameters were chosen to be the converged results from many
test fits where a random point in the allowed range of the
multidimensional parameter space was used as the start for the
minimization. After a call of minimization by the method of MIGRAD,
the statistic error estimation was done by a MINOS error analysis. The
returned symmetric parabolic error was reported.

Several procedures were checked to ensure the fit quality. At first,
the fit status returned by MINUIT was required to have a normal
convergence. The global correlation coefficient of each parameter
should be greater than zero and less than 0.99, to avoid a wrong
estimation of the statistic error for the case of uncorrelated or
strongly correlated parameters. The typical value of the global
correlation coefficient was distributed between 0.2 and 0.6. Second,
the fit parameter for the number of events must be consistent with the
input statistics, an advantage from the use of an extended maximum
likelihood fit. Finally, reasonable agreement between the
one-dimensional angular distributions of $\cos\theta$, $\varphi$,
$\Phi$, and $\varphi-\Phi$ from the input events and the fitted values
of $\hat{\rho}$, was checked. The $\chi^2$ value per degree of freedom
from a normalization fit is required to be in the range of 1.0-3.0. In
general, a fit for a given kinematic bin associated with small
statistics was likely to fail in this quality check.

The spin-density matrix elements do not depend on the beam
normalization which is typically the main systematic uncertainty of
cross-section measurements at LEPS. We evaluated the systematic
uncertainty in the following way. Ensembles of MC events were
generated with specific sets of spin-density matrix elements being the
same as those found in the real data. Those MC events were then
filtered by the experimental detection efficiency. Subsequently, those
events with statistics similar to the real data were analyzed for the
determination of spin-density matrix elements $\hat{\rho}$. The mean
of the distribution of obtained $\hat{\rho}$ for many trials, the
so-called pull distribution, was determined. The deviation of the
mean from the generated value of $\hat{\rho}$ for MC events
contributed to the estimation of the systematic uncertainty.

Another source came from the background reactions [e.g., nonresonant
$K^+K^-$ and $\Lambda(1520)$ production]. This bias was estimated by
comparing the results in the signal region [$|{\rm M}(K^+K^-)-{\rm
M}_{\phi}|<$0.01 GeV/c$^2$] and the sideband region [$0.01<|{\rm
M}(K^+K^-)-{\rm M}_{\phi}|<$0.02 GeV/c$^2$], by taking into account
different signal-to-background ratios in these two regions . In
general, the signal-to-background ratio in the defined $\phi$-event
region is good enough as shown in Figs.~\ref{fig:spectrum1_LH2}(a) and
\ref{fig:spectrum1_LH2}(b) and Figs.~\ref{fig:spectrum1_LD2}(a) and
\ref{fig:spectrum1_LD2}(b). Thus this bias was found to be small (of
the order of 0.01$-$0.05 except at regions of larger $|t|$). Since the
statistics in the sideband region was not always enough for a reliable
fit, we conservatively included this bias in the systematic error,
rather than applying the corresponding correction to the results.

For the $\phi$-meson production from deuterium, there are additional
sources of systematic error: the off-shell effect for the incoherent
process in the MC simulation and the division point in the missing
mass regions for
disentanglement~\cite{LEPS_phiD_co,LEPS_phiD_inco}. The choice of the
division point $\rm MM_{\rm div}$ would affect the contents of
coherent and incoherent events on the two separated regions. Four
choices of $\rm MM_{\rm div}$, 1.875, 1.88, 1.89 and 1.90 GeV/c$^2$,
were used for evaluating the systematic uncertainty.

\section {Results}
\label{sec:results}

We measured the spin-density matrix elements of $\phi$-meson
photoproduction in the region of $1.57<E_{\gamma}<2.37$ GeV and
$|\tilde{t}|<0.2$ GeV$^2$/c$^2$. We define $\tilde{t}$ as
$t-t_{min}^{p}$ or $t-t_{min}^{d}$, which is the squared four-momentum
transfer $t$ subtracted by its minimum value for the corresponding
photon energy bin, under the assumption that a proton or deuteron is
at rest. The binning sizes were 0.2 GeV for \Eg\ and 0.05
GeV$^2$/c$^2$ for $\tilde{t}$. The real or imaginary parts of
spin-density matrix elements \sdmij\ for the \phipro, \phidpn, and
\phidd\ reactions were determined by the method as described in
Secs.~\ref{sec:method} and ~\ref{sec:exp}.

\subsection{Production from free protons: \phipro}
\label{subsec:results_LH2}

\begin{figure}[htbp]
\includegraphics[width=\factor\textwidth]{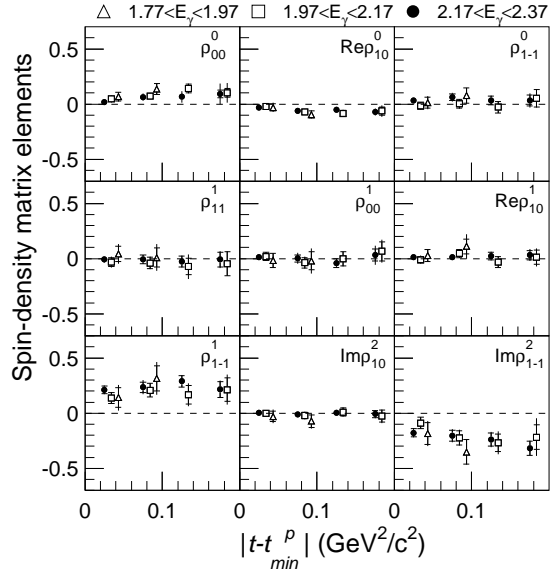}
\caption{Spin-density matrix elements for the \phipro\ reaction in the
helicity system as a function of $|t-t_{min}^p|$ in various \Eg\
regions.  The vertical bars are for statistical error only.}
\label{fig_sdm_LH2_H}
\end{figure}

\begin{figure}[htbp]
\includegraphics[width=\factor\textwidth]{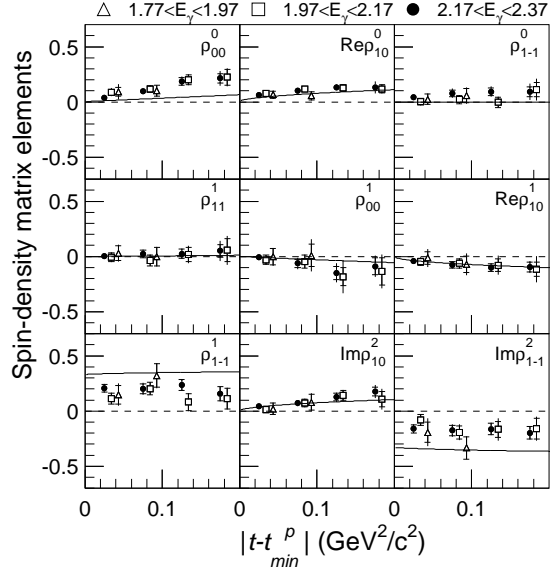}
\caption{Spin-density matrix elements for the \phipro\ reaction in the
  Gottfried-Jackson system as a function of $|t-t_{min}^p|$ in various
  \Eg\ regions. The vertical bars are for statistical error only. The
  solid lines are the theoretical predictions of $\hat{\rho}$ for the
  \phipro\ reaction at $\Eg=2$ GeV~\cite{Titov_iso2}.}
\label{fig_sdm_LH2_GJ}
\end{figure}

\begin{figure}[htbp]
\includegraphics[width=\factor\textwidth]{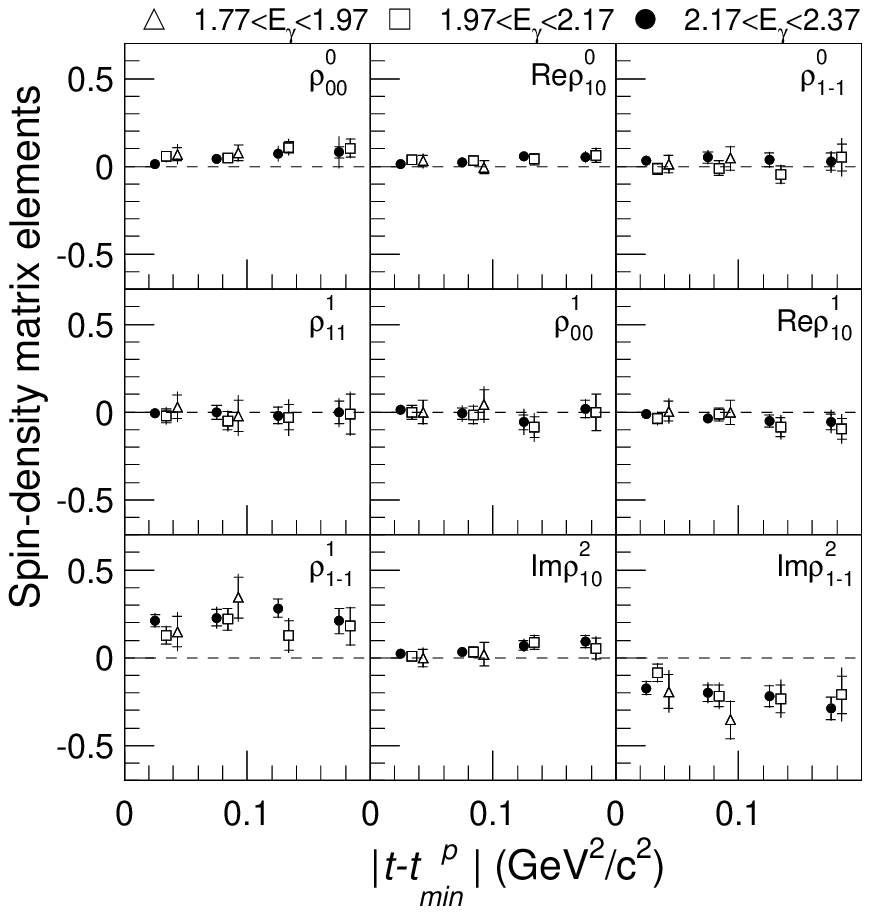}
\caption{Spin-density matrix elements for the \phipro\ reaction in the
  Adair system as a function of $|t-t_{min}^p|$ in various \Eg\
  regions. The vertical bars are for statistical error only.}
\label{fig_sdm_LH2_A}
\end{figure}

The spin-density matrix elements of the \phipro\ reaction in the
helicity, Gottfried-Jackson, and Adair systems are shown in
Figs.~\ref{fig_sdm_LH2_H}, \ref{fig_sdm_LH2_GJ} and
\ref{fig_sdm_LH2_A}, respectively. Within the error there is no strong
energy dependence in the measured region of $1.77<\Eg<2.37$ GeV. Clear
nonzero values for \rhog\ and \rhoi\ are seen in all three systems,
and $\rhog \approx 0.2$. According to Eq.~(\ref{eq:nunratio}), a sizable
30\% contribution of unnatural-parity exchange processes, other than
the the natural-parity Pomeron exchange, is observed.

In the helicity system, small but nonzero positive values for \rhoa\
and negative ones for \rhob~are observed. The \rhoa\ (\rhob) are of
positive (negative) values. The finiteness of \rhoa\ indicates the
presence of amplitudes violating $s$-channel-helicity in the $\phi$
production.

In the Gottfried-Jackson system, \rhoa\ seems to be larger and \rhob\
turns to be positive, compared to the results in the helicity
system. The increase in the magnitude of \rhoa\ suggests a stronger
violation of helicity conservation in the $t$ channel. Specifically,
those spin-density matrix elements reflecting helicity-nonconserving
amplitudes such as \rhoa, \rhoc, \rhoe, \rhof, and \rhoh become
clearly nonzero at large $\tilde{t}$. Figure~\ref{fig_sdm_LH2_GJ}
shows a comparison with the prediction from a model based on the
dominance of the Donnachie-Landshoff Pomeron plus
($\pi,\eta$)-exchange channels~\cite{Titov_iso2} at $\Eg=2$ GeV. This
model correctly predicts the sign of each matrix element. However, the
${t}$ dependence for \rhoa, \rhoc, and \rhoe\ is not correctly
predicted. Also the absolute scale of \rhog\ and \rhoi\ are
overpredicted, which means that the contribution from the
unnatural-parity exchange processes is not sufficient in this model.

The $\hat{\rho}$ values in the Adair system are rather similar to those in
the helicity system except for an opposite sign for \rhob\ and \rhoh.

\subsection{Incoherent production from deuterons: \phidpn}
\label{subsec:results_LD2inco}

\begin{figure}[htbp]
\includegraphics[width=\factor\textwidth]{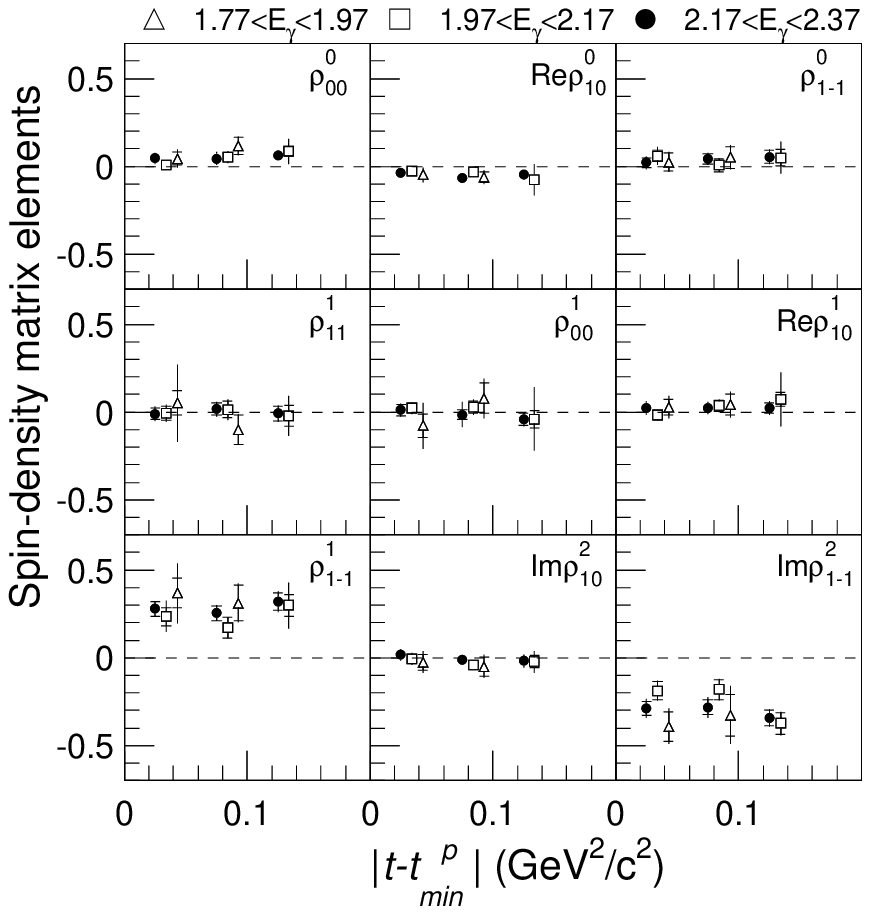}
\caption{Spin-density matrix elements for the \phidpn\ reaction in the
  helicity system as a function of \Eg\ and $|t-t_{min}^p|$.  The
  vertical bars are for statistical error only.}
\label{fig_sdm_LD2inco_H}
\end{figure}

\begin{figure}[htbp]
\includegraphics[width=\factor\textwidth]{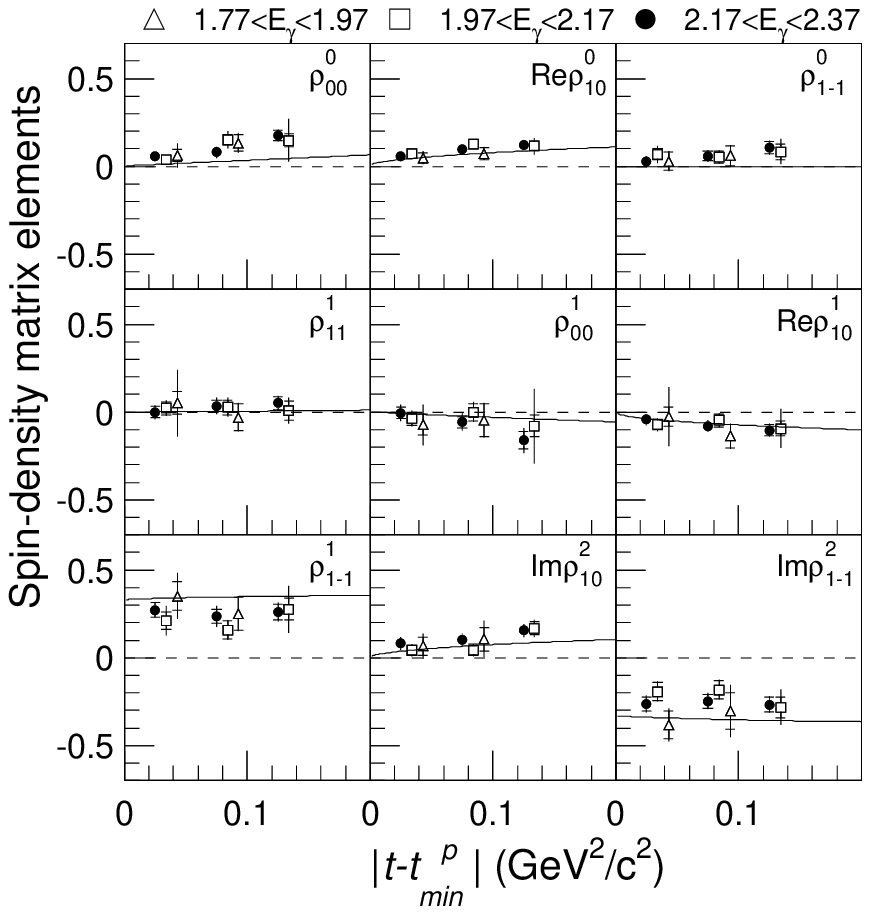}
\caption{Spin-density matrix elements for the \phidpn\ reaction in the
  Gottfried-Jackson system as a function of \Eg\ and $|t-t_{min}^p|$.
  The vertical bars are for statistical error only.  The solid lines
  are the theoretical predictions of $\hat{\rho}$ for the \phipro\ reaction
  at $\Eg=2$ GeV~\cite{Titov_iso2}.}
\label{fig_sdm_LD2inco_GJ}
\end{figure}

\begin{figure}[htbp]
\includegraphics[width=\factor\textwidth]{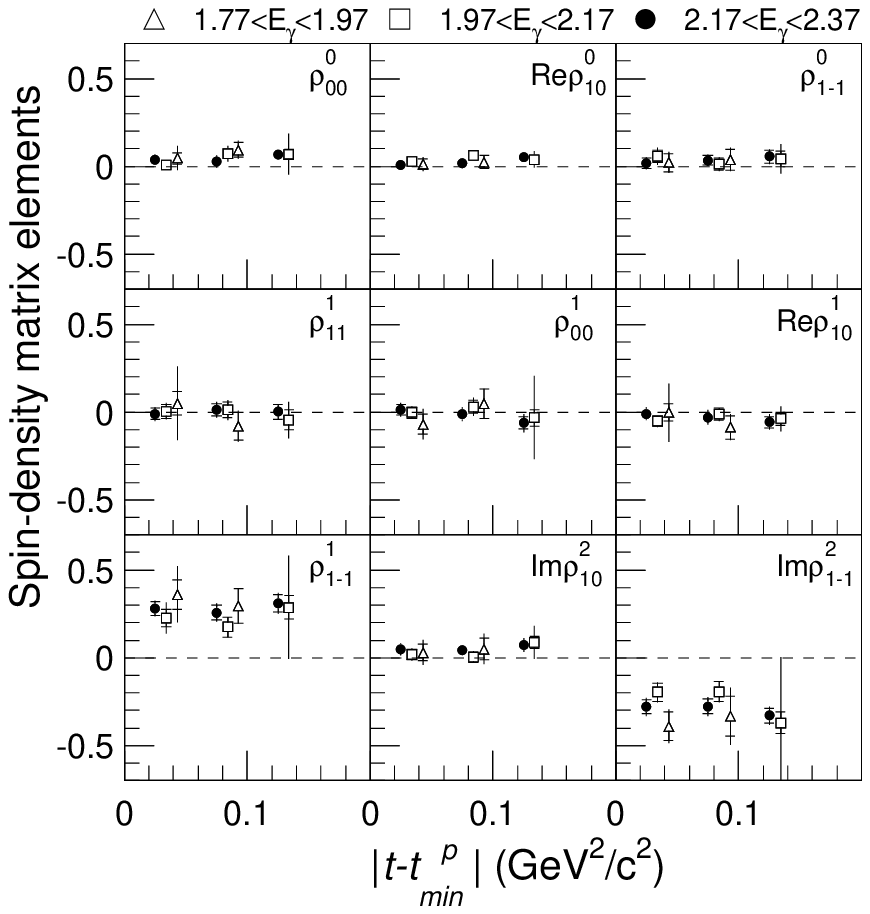}
\caption{Spin-density matrix elements for the \phidpn\ reaction in the
  Adair system as a function of \Eg\ and $|t-t_{min}^p|$.  The vertical
  bars are for statistical error only.}
\label{fig_sdm_LD2inco_A}
\end{figure}

The spin-density matrix elements for the incoherent production from
deuterons in three angular systems are shown in
Figs.~\ref{fig_sdm_LD2inco_H}, \ref{fig_sdm_LD2inco_GJ} and
\ref{fig_sdm_LD2inco_A}, respectively. The energy dependence is also
insignificant. Because of statistical constraint, the measured range
of $\tilde{t}$ is limited to $|\tilde{t}|<0.15$. In general, the
results are quite similar to those of the production from free protons
except that the absolute values of \rhog\ and \rhoi\ are slightly
larger ($\rhog \approx 0.25$). This suggests that the contribution
from unnatural-parity exchange processes is reduced in the production
from neutrons. It could be interpreted as a destructive interference
effect among the unnatural-parity ($\pi,\eta$)-exchange processes in
the $\phi$-meson production from the
neutron~\cite{Titov_iso,Titov_iso2}.

\subsection{Coherent production from deuterons: \phidd}
\label{subsec:results_LD2co}

\begin{figure}[htbp]
\includegraphics[width=\factor\textwidth]{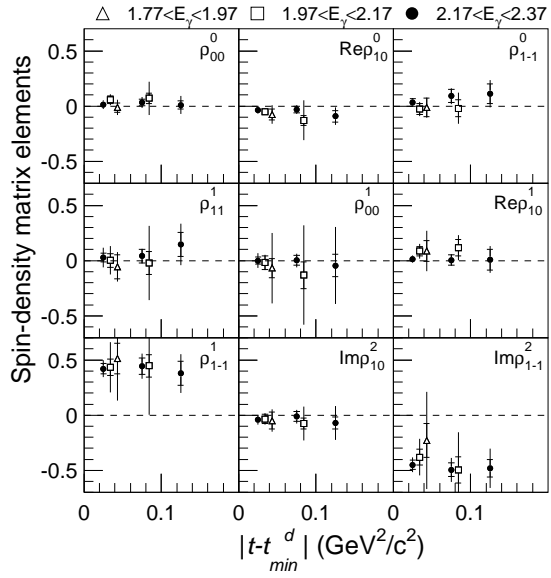}
\caption{Spin-density matrix elements for the \phidd\ reaction in the
  helicity system as a function of \Eg\ and $|t-t_{min}^d|$.  The
  vertical bars are for statistical error only.}
\label{fig_sdm_LD2co_H}
\end{figure}

\begin{figure}[htbp]
\includegraphics[width=\factor\textwidth]{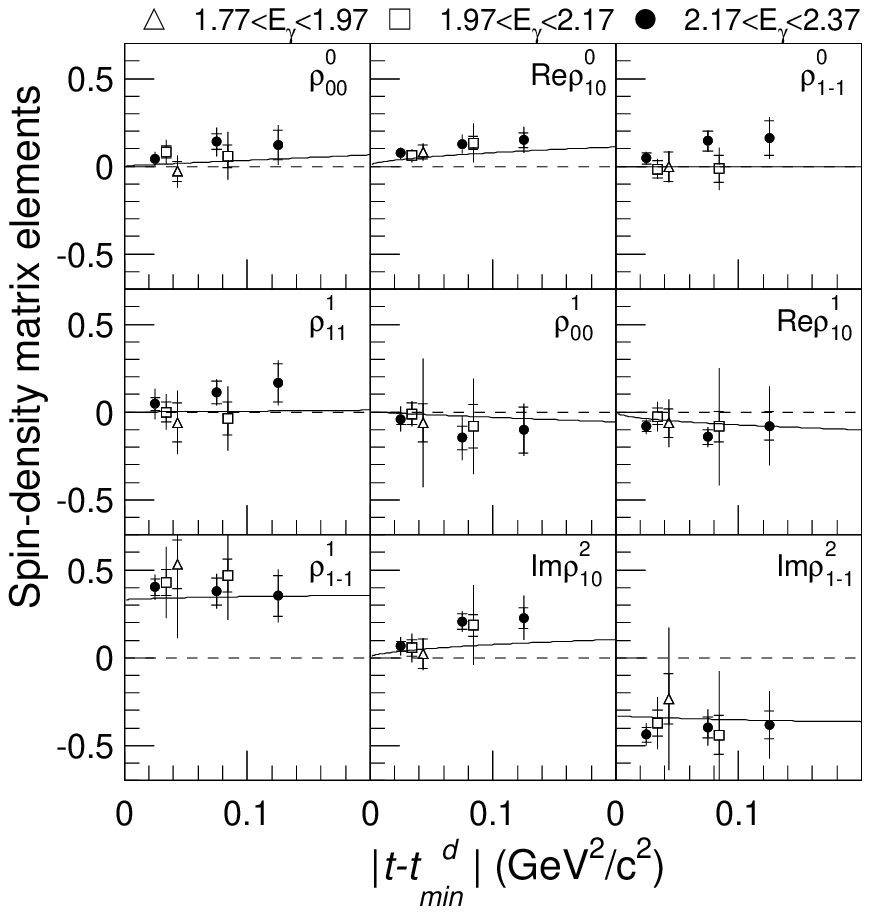}
\caption{Spin-density matrix elements for the \phidd\ reaction in the
  Gottfried-Jackson system as a function of \Eg\ and $|t-t_{min}^d|$.
  The vertical bars are for statistical error only. The solid lines
  are the theoretical predictions of $\hat{\rho}$ for the \phipro\ reaction
  at $\Eg=2$ GeV~\cite{Titov_iso2}.}
\label{fig_sdm_LD2co_GJ}
\end{figure}

\begin{figure}[htbp]
\includegraphics[width=\factor\textwidth]{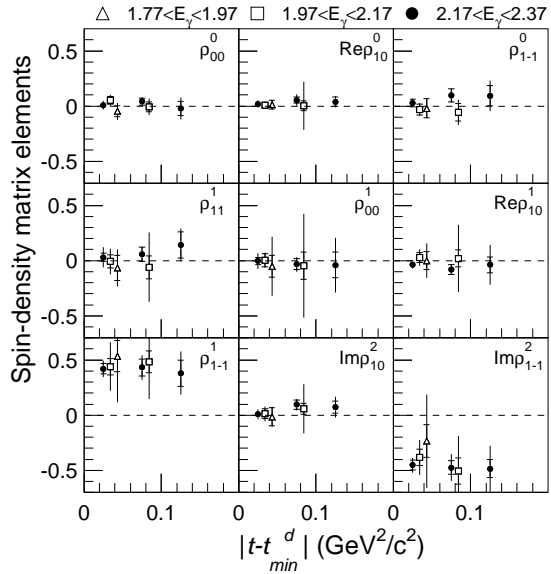}
\caption{Spin-density matrix elements for the \phidd\ reaction in the
  Adair system as a function of \Eg\ and $|t-t_{min}^d|$.  The vertical
  bars are for statistical error only.}
\label{fig_sdm_LD2co_A}
\end{figure}

Figures~\ref{fig_sdm_LD2co_H}, \ref{fig_sdm_LD2co_GJ} and
\ref{fig_sdm_LD2co_A} display the spin-density matrix elements for the
coherent production from deuterium in three angular systems. A
distinct feature is a strong increase of the absolute values of \rhog\
and \rhoi, and $\rhog \approx 0.45$, compared to what is observed in
the reactions of \phipro\ and \phidpn. This again suggests that the
contribution from the unnatural-parity exchange processes is
significantly reduced. It could be understood as a result of the
forbidden coupling of the isovector $\pi$ exchange with the isoscalar
deuteron target. All the other components in $\hat{\rho}$ are similar
to those in the production from nucleons. In the helicity system the
smallness of \rhoa\ (less than 10\%) in the small-$|t|$ region is the
same as what was observed in the region of large $|t|$ by
CLAS~\cite{CLAS_phiD_co}.

\section {Discussion}
\label{sec:discussion}

The finiteness of the spin-density matrix elements \rhoa, \rhoc,
\rhoe, \rhof, and \rhoh is clearly observed in the Gottfried-Jackson
system. This suggests the presence of helicity-nonconserving effects
in the $t$-channel exchange processes for $\phi$-meson
photoproduction, (e.g., two-gluon exchange as mentioned
before~\cite{Laget}). Relatively, the helicity-nonconserving effects
become less in the helicity ($s$-channel) and Adair systems.

Previously, we reported a nonzero value of \rhoc\ ($0.12 \pm 0.03$)
for the \phipro\ reaction in the region of $|t-t_{min}^{p}|<0.2$
GeV$^2$/c$^2$ at $\Eg=1.77-1.97$ GeV~\cite{LEPS_phip}, while \rhoc\
became less (about $0.04\pm0.02$) at $\Eg=1.97-2.17$ GeV. The
measurement was done with a fit on the one-dimensional azimuthal angle
distribution $W(\varphi)$ in the Gottfried-Jackson system. In
Refs.~\cite{Titov_coherent,Titov_coherent2}, the authors argued that
\rhoc, which could only come from a spin-orbit interaction, must be
close to zero near threshold and should monotonically increase with
photon energy. However, the results were obtained in the region of
$|t-t_{min}^{p}|<0.2$ GeV$^2$/c$^2$, instead of at
$t=t_{min}^{p}$. The effect of $\tilde{t}$ dependence shown in
Fig.~\ref{fig_sdm_LH2_GJ} should be taken into account.

\begin{figure}[htbp]
\includegraphics[width=\factor\textwidth]{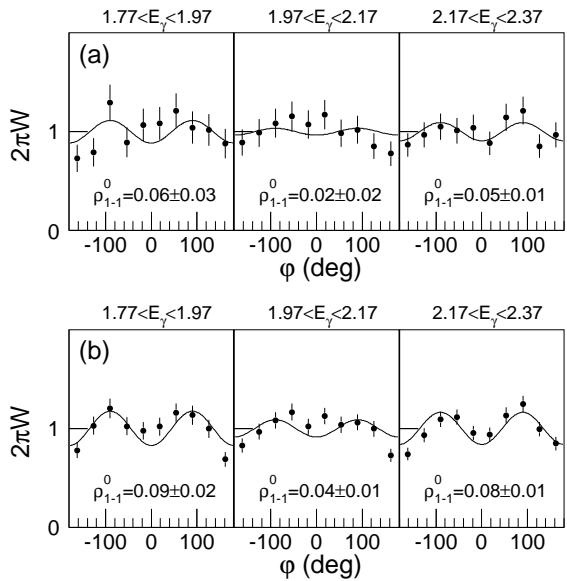}
\caption{(a) Azimuthal angle distribution $W(\varphi)$ for \phipro\
at $|t-t_{min}^{p}|<0.05$ GeV$^2$/c$^2$ in the Gottfried-Jackson
system. The solid curves are the fit to the data. (b) Same as (a) but
at at $|t-t_{min}^{p}|<0.2$ GeV$^2$/c$^2$.}
\label{fig_lh2_azmuthal}
\end{figure}

Using the current data set from the hydrogen runs, we repeated the
same analysis. The results are shown in
Fig.~\ref{fig_lh2_azmuthal}. Figures~\ref{fig_lh2_azmuthal}(a) and
\ref{fig_lh2_azmuthal}(b) display the $W(\varphi)$ distributions and
the fitted \rhoc\ for the \phipro\ reaction in the regions of
$|t-t_{min}^{p}|<0.05$ GeV$^2$/c$^2$ and $|t-t_{min}^{p}|<0.2$
GeV$^2$/c$^2$ at three \Eg\ bins. Obviously there is some $t$
dependence. The current measurement at $|t-t_{min}^{p}|<0.2$
GeV$^2$/c$^2$ is consistent with the previous results. At the smallest
$|\tilde{t}|$ bin, \rhoc\ is about 0.05, which is consistent with the
results shown in Fig.~\ref{fig_sdm_LH2_GJ}. Within the error, we could
not draw any conclusion on the monotonic increase of \rhoc\ with the
photon energy but at least there is no large jump in the energy
dependence. It is interesting to note that the \rhoc\ value at
$\Eg=1.97-2.17$ GeV, where the peak of the bump structure in the
$\phi$-meson production cross section appears, seems systematically
smaller than those in the neighboring energy bins.

The measurements of the diffractive $\phi$ production at low energies
are qualitatively consistent with the description of natural-parity
Pomeron and unnatural-parity pseudoscalar exchanges in the $t$
channel~\cite{Titov_spinamp}. The latter term includes the
($\phi,\eta$) exchange and accounts for about 30\% of the production
cross section in the \phipro\ reaction. In the coherent production
process, the isovector $\pi$ exchange is eliminated, and the
natural-parity exchange becomes quite dominant. This observation
reflects the fact that the $\eta$-exchange channel is relatively
small. A slight increase in the contribution for natural-parity
exchange is seen in the incoherent production, being consistent with
the theoretical prediction, if one takes into account the destructive
effect of $\pi$- and $\eta$-exchanges in the $\gamma n \to \phi n$
reaction~\cite{Titov_iso,Titov_iso2}.

A bump structure in the differential cross section at forward angles
is observed at $\Eg=2$ GeV in both the production from free protons
and the incoherent production from deuterons. The origin of this
structure should be common for production from both protons and
neutrons. Thus it is unlikely that the interference effect from the
($\pi,\eta$) exchange is responsible~\cite{Titov_coherent2}. Judging
from the smallness of $\hat{\rho}$ except \rhog\ and \rhoi\ in the
helicity frame, we see that the helicity-nonconserving amplitudes is
rather limited for the $s$ channel in the very forward
direction. There is also no significant variation in the spin-density
matrix elements across the bump region. Therefore we conclude that
this bump structure is also unlikely to be caused by any nucleon
resonance states containing large $s\bar{s}$
content~\cite{Hosaka_phi,Kiswandhi_phi} unless there are some
complicated interference effects.

\section{Summary}
\label{sec:summary}

In summary, we presented the measurement of decay angular
distributions of $\phi$ mesons produced from protons and deuterons
with linearly polarized photons at very forward angles at
$\Eg=1.77-2.37$ GeV. Nine spin-density matrix elements representing the
angular distributions in the \KK\ decay mode are obtained
simultaneously by the extended maximum likelihood fit in three
different coordinate frames. There is no strong energy dependence
observed in the measured energy region, although some $t$ dependence
is found. Small but finite helicity-nonconserving effects are seen in
the $t$ channel but they become less significant in the $s$ channel.

Based on the measurement of \rhog\ and \rhoh, unnatural-parity
exchange processes are no longer negligible near threshold, even
though the natural-parity Pomeron exchange process still
dominates. The unnatural-parity processes are significantly reduced in
the coherent production from deuterium. This fact can be interpreted
to appear as the result of the suppression of isovector
$\pi$ exchange. The averaged contribution of the unnatural-parity
processes in incoherent production decreases slightly and could be due
to the destructive interference of the ($\pi,\eta$)-exchange processes
in production from neutrons.

A theoretical model based on the dominance of the Donnachie-Landshoff
Pomeron plus ($\pi,\eta$)-exchange channels~\cite{Titov_iso2} gives a
reasonable prediction at $\Eg=2$ GeV but obviously there is room for
quantitative improvement. New and comprehensive information on spin
observables available from this work should help to differentiate the
theoretical models. Hopefully, this will lead to a better picture of the
$\phi$-meson photoproduction at low energies, especially to shed light
on whether any exotic channel is necessary to account for the bump
structure at $\Eg \sim 2$ GeV.

A manifestation of an unnatural-parity exchange component in the
photoproduction of $\phi$ mesons near threshold regions was already
observed. New programs at LEPS and CLAS aim to perform
measurements at $\Eg=2.5-3.5$ GeV, and to study double polarization
observables with polarized proton and deuterium targets. Such
experimental efforts shall bring further understanding of the
appearance of bump structure in the $\phi$-meson photoproduction near
threshold.

\begin{acknowledgments}
The authors thank the SPring-8 staff for their great help during the
operation of the LEPS experiment. This research was supported in part
by the Ministry of Education, Science, Sports and Culture of Japan, the National Science Council of Republic of China (Taiwan), Korea
Research Foundation Grant No. 2009-0089525, and National Science
Foundation Grant No. PHY-0653454.
\end{acknowledgments}



\begin{thebibliography}{99}

\bibitem{LEPS_phip} T. Mibe {\it et al.} (LEPS Collaboration),
  Phys. Rev. Lett. {\bf 95}, 182001 (2005).

\bibitem{LEPS_phiD_co} W.C. Chang {\it et al.} (LEPS Collaboration),
  Phys. Lett. B {\bf 658}, 209 (2008).

\bibitem{LEPS_phiD_inco} W.C. Chang {\it et al.} (LEPS Collaboration),
  Phys. Lett. B {\bf 684}, 6 (2010).

\bibitem{LEPS_phiA_inco} T. Ishikawa {\it et al.} (LEPS
  Collaboration), Phys. Lett. B {\bf 608}, 215 (2005)

\bibitem{CLAS_phip} E. Anciant {\it et al.} (CLAS Collaboration),
  Phys. Rev. Lett. {\bf 85}, 4682 (2000).

\bibitem{CLAS_phiD_co} T. Mibe {\it et al.} (CLAS Collaboration),
  Phys. Rev. C {\bf 76}, 052202 (2007).

\bibitem{CLAS_phiD_inco} X. Qian {\it et al.} (CLAS Collaboration),
  Phys. Lett. B {\bf 680}, 417 (2009).

\bibitem{SAPHIR_phip} J. Barth {\it et al.} (SAPHIR Collaboration),
  Eur. Phys. J. A {\bf 17}, 269 (2003).

\bibitem{Schilling} K. Schilling, P. Seyboth, and G. Wolf,
  Nucl. Phys. {\bf B15}, 397 (1970).

\bibitem{Titov_iso} A.I. Titov, T.S.-H. Lee, and H. Toki, Phys. Rev. C
  {\bf 59}, R2993 (1999).

\bibitem{Titov_iso2} A.I. Titov, T.S.-H. Lee, H. Toki, and
  O. Streltsova, Phys. Rev. C {\bf 60}, 035205 (1999).

\bibitem{Titov_spinamp} A.I. Titov and T.S.-H. Lee, Phys. Rev. C {\bf
  67}, 065205 (2003).

\bibitem{Titov_coherent} A. I. Titov, M. Fujiwara, and T. S.-H. Lee
  Phys. Rev. C {\bf 66}, 022202 (2002).

\bibitem{Titov_coherent2} A.I. Titov and B. K\"{a}mpfer,
  Phys. Rev. C {\bf 76}, 035202 (2007).

\bibitem{phi_Hpalpern} H. J. Halpern {\it et al.}, Phys. Rev. Lett. {\bf 29},
  1425 (1972).

\bibitem{phi_Ballam} J. Ballam {\it et al.}, Phys. Rev. D {\bf 7},
3150 (1973).

\bibitem{phi_Atkinson} M. Atkinson {\it et al.}, Z. Phys. {\bf C 27},
233 (1985).

\bibitem{Hosaka_phi} S. Ozaki, A. Hosaka, H. Nagahiro, and
O. Scholten, Phys. Rev. C {\bf 80}, 035201 (2009); {\bf 81}, 059901(E)
(2010).

\bibitem{Kiswandhi_phi} A. Kiswandhi, J.J. Xie, and S.N. Yang,
Phys. Lett. B {\bf 691}, 214 (2010).

\bibitem{phi_Ballam2} J. Ballam {\it et al.}, Phys. Rev. D {\bf 5},
545 (1972).

\bibitem{Titov_ssknockout} A.I. Titov, Y. Oh, S.N. Yang, and T. Morii,
Phys. Rev. C {\bf 58}, 2429 (1998)

\bibitem{Zhao} Q. Zhao, B. Saghai, and J. S. Al-Khalili, Phys. Lett. B
{\bf 509}, 231 (2001).

\bibitem{Laget} J.-M. Laget, Phys. Lett. B {\bf 489}, 313 (2000).

\bibitem{MINUIT} F. James, CERN Applications Software Group, CERN
Program Library Long Writeup D506, 1998.

\bibitem{Obasis_Chung} S. U. Chung {\it et al.}, Phys. Rev. Lett. {\bf
40}, 355 (1978); S. U. Chung, Phys. Rev. D {\bf 56}, 7299 (1997).

\bibitem{EML_Lyons} L. Lyons, W. M. Allison, and J. P. Comellas,
Nucl. Instrum. Methods Phys. Res., Sect. A {\bf 245}, 530 (1986);
R. Barlow, {\it ibid.}, Sect. A {\bf 297}, 496 (1990).

\bibitem{LEPS} M. Sumihama {\it et al.} (LEPS Collaboration),
Phys. Rev. C {\bf 73}, 035214 (2006).

\end{thebibliography}
\end{document}